# Methods for Detecting Paraphrase Plagiarism

Victor U Thompson and Chris Bowerman

Department of Computer Science

University of Sunderland, SR1 3SD

{Victor.thompson@research, chris.bowerman}@sunderland.ac.uk

*Abstract:* Paraphrase plagiarism is one of the difficult challenges facing plagiarism detection systems. Paraphrasing occur when texts are lexically or syntactically altered to look different, but retain their original meaning. Most plagiarism detection systems (many of which are commercial based) are designed to detect word co-occurrences and light modifications, but are unable to detect severe semantic and structural alterations such as what is seen in many academic documents. Hence many paraphrase plagiarism cases go undetected. In this paper, we approached the problem of paraphrase plagiarism by proposing methods for detecting the most common techniques (phenomena) used in paraphrasing texts (namely; lexical substitution, insertion/deletion and word and phrase reordering), and combined the methods into a paraphrase detection model. We evaluated our proposed methods and model on collections containing paraphrase texts. Experimental results show significant improvement in performance when the methods were combined (the proposed model) as opposed to running them individually. The results also show that the proposed paraphrase detection model outperformed a standard baseline (based on greedy string tilling), and previous studies.

## 1  Introduction

While significant progress has been made in recent years to tackle plagiarism using automated systems, more still needs to be done. Most plagiarism detection software are not well equipped to deal with paraphrase plagiarism (Barrón-Cedeño, Vila, Martí, and Rosso 2013), translational and idea plagiarism (Oberreuter and VeláSquez, 2013; Meuschke and Gipp, 2013). Paraphrase plagiarism is the most prevalent of these challenges, because it is relatively easier to carry out, and often used in conjunction with other types of plagiarism, such as in translation plagiarism where a piece of texts is translated into a different language and paraphrased to obfuscate plagiarism (Barrón-Cedeño, Gupta, and Rosso, 2013). Paraphrasing occur when texts are lexically or syntactically modified (Clough and Stevenson, 2011) to look differently from their sources, but retain the same meaning. Paraphrasing itself is legal when done properly such as in Journalistic text reuse (Clough, Gaizauskas, and Piao, 2002), but when texts are modified and used without properly acknowledging the sources, it is plagiarism. Attempts made in previous studies to address the problem of paraphrase plagiarism detection were not really successful. This is evident from the results obtained in the annual competition on plagiarism detection organised by Pan (Potthast et al., 2010; 2011; 2012; 2014). In the Pan2010 competition, only a third of the simulated paraphrase cases were detected correctly (Barrón-Cedeño et al., 2013; Meuschke and Gipp, 2013), a similar result was obtained the following year. These results reveal weaknesses in conventional plagiarism detection systems when faced with paraphrase plagiarism.

Traditional information retrieval (IR) methods such as the vector space model (VSM) do not perform well on paraphrase plagiarism due to vocabulary mismatch that occur when words are replaced with their synonyms. Semantic similarity measurement is required to effectively detect lexical substitutions (synonym replacements), and current techniques are based on query expansion using lexical resources such as WordNet (Chong and Specia, 2011; Nawab et al., 2012; 2016), most of which are limited in vocabulary size, restricted to only certain parts of speech, and not effective in identifying the right sense of a word in context. Hence, a semantic similarity measurement technique that accesses a much larger vocabulary base, and uses word sense disambiguation (WSD) to increase the chances of detecting replaced words is required. Besides the problem of synonym replacement, studies on existing paraphrased corpora revealed that lexical substitutions (e.g. synonym replacement); insertions/deletions (Barrón-Cedeño et al., 2013) and word reordering are the most common paraphrase techniques used in modifying texts. An ideal paraphrase detection tool should therefore be able to detect these common paraphrase techniques, as most real-word cases of paraphrase plagiarism are likely to fall under one or more of these categories.

In this paper we approached the problem of paraphrase plagiarism by proposing methods for detecting the most common techniques used in paraphrasing texts, namely; synonym replacement (lexical substitutions), word reordering (syntactic alterations) and deletions/insertions (edit operations). We combined the methods into a paraphrase detection model and evaluated and generalised the methods and model on paraphrased corpora, and made comparisons with a standard baseline and previous studies. The rest of this paper is divided into the following sections; the contributions of this work to knowledge, a review of the relevant literature, proposed paraphrased detection methods, experiments and evaluations, and conclusion.

## 2 Contributions

This article makes the following contributions to knowledge;

1. We proposed a model for detecting paraphrase plagiarism that combines measurements across semantic, syntactic and insert/delete intertextual similarity dimensions when comparing texts.
2. We demonstrated empirically that the best approach for detecting paraphrase plagiarism is to combine methods specifically designed to detect the most common techniques used in paraphrasing texts, using suitable machine learning algorithms.

# 3 Relevant Literature

Common approaches used in the literature to detect paraphrases or obfuscation plagiarism typically involve the application of semantic or syntactic (structural) similarity measurement, or a combination of methods.

Semantic similarity measurement involves comparing texts for similarity in meaning. In plagiarism detection, semantic similarity measurement is often applied using query expansion; the process involves searching for plagiarised words that may have been replaced with their synonyms using lexical databases such as WordNet (Chong and Specia, 2011; Nawab, Stevenson, and Clough, (2012, 2016)). One major challenge with query expansion is that words may mean different things in different contexts (polysemy; 'bat' is a bird and also a sport equipment), some degree of word sense disambiguation (WSD) is therefore required to determine the right sense of a word in context (McInnes and Pedersen, 2013; Jurafsky and Martin, 2015). Word pair similarity measurement is another semantic similarity technique that is not common in plagiarism detection. Semantic similarity between words can be estimated using concepts in WordNet taxonomy, or by comparing word vectors using word embedding architectures. Several tools have been proposed for comparing word pairs using WordNet concepts (Pedersen, Patwardhan, and Michelizzi, 2004; McInnes and Pedersen, 2013; Vrbanec and Meštrović, 2017). These tools compute similarity between words using the path length (depth) of their least common subsumer (LCS) in a WordNet taxonomy, or the information content (IC=-$log\ p$) of their LCS. The LCS of a pair of words is a concept that subsumes the pair. A study by McInnes and Pedersen, (2013) revealed that IC based measures are more effective than those based on path length. In similar studies where word pair similarity measures (Resnik IC, Wu and Palmer, Lin etc) were compared (Mihalcea, Corley, and Strapparava, 2006; Pesaranghader, Matwin, Sokolova and Beiko, 2015), the Resnik IC emerged best performing. The Resnik IC measures the similarity between two concepts using the IC of their LCS; Resnik $(c_1, c_2) = -\log(P(LCS(c_1, c_2)))$. Words that have similar concepts have certain degree of similarity, when similarity scores from such words in a pair of sentences are aggregated and normalised by the length of the sentences, the resultant score is an estimation of the semantic similarity of the sentences. Word embedding on the other hand involves modelling words and their contexts into vectors of real numbers, and measuring the similarity between words based on vector comparison. The idea here is that words that appear often in similar contexts are similar (distributional hypothesis). One of the most common word embeddings architecture is the word2vec model proposed by Mikolov, Corrado, and Dean, (2013) where words are transformed into vectors using either a continuous bag of word (CBOW) or skip-gram model, and trained to predict similar words using deep neural network architecture. Ferrero, Agnes, Besacier, and Schwab,(2017) applied word2vec to detect cross lingual plagiarism by using bilingual embeddings based on the CBOW model to generate word vectors for words in a pair of sentences written in different languages, and comparing the sentences for similarity using the cosine measure, sentence pairs with similarity above predefined threshold are considered potentially plagiarised. Konopik, Prazák, Steinberger, and Brychcín (2016) applied modified lexical semantic vector with word2vec model to determine the semantic similarity of text pairs. The modified lexical semantic vector is a variant of the word semantic vector proposed in Li, McLean, Bandar, O'shea, and Crockett (2006), but uses word2vec to estimate the similarity between words instead of an information content measure.

Structural or syntactic methods use the syntactic features of texts such as word order, stopword patterns (Stamatatos, 2011), part of speech (POS) similarity (Alzahrani, Salim, and Abraham, 2012) to detect paraphrase texts. In many cases of plagiarism, words are often replaced with other words of the same POS; also known as POS preserving (Barrón-Cedeño, Potthast, Rosso, Stein, and Eiselt, 2010; Chong, Specia, and Mitkov, 2010). Similar texts should therefore have similar POS patterns even when they have different lexicons. In some studies syntactic similarity measurement is applied by restricting document comparison to only certain parts of speech, usually nouns, adjectives, verbs and adverbs (Alzahrani et al., 2012; Williams, Chen, Choudhury, and Giles, 2013). This is because words of such parts of speech are more reflective of plagiarism, but this technique is prone to information loss. Index alignment between pairs of sentences has also been proposed for measuring the syntactic similarity between texts. Abdi, Idris, Alguliyev, and Aliguliyev (2015) applied a method originally proposed in Li et al., (2006) to detect reordering in paraphrased sentences by creating vectors using the indices of similar words in a pair of sentences, and computing similarity using the normalised differences of the vectors. Similarity in the patterns of stopword n-grams between texts have also been proposed for detecting high obfuscation plagiarism (Stamatatos, 2011).

Other approaches used for paraphrase detection combines several methods. Zesch and Gurevych (2012) combined content, structural and stylistic based similarity measures to detect paraphrase plagiarism. Content based measures use textual features such as word co-occurrences to measure intertextual similarity (example: cosine similarity measure), while stylistic method looks for similarity in writing style (example: vocabulary richness, term frequency (TF)). Combinations from Zesch and Gurevych (2012) experiments outperformed single methods on paraphrase detection. Pertile, Moreira, and Rosso (2015) combined content and citation based methods and obtained encouraging results. Citation method measures the similarity between citations in a pair of documents (Gipp, 2014), and works best when combined with well-established methods (i.e. string matching methods). One content based method that was outstanding when testing the P4P paraphrase corpus is the Greedy String Tilling (GST) (Nawab et al., 2010; Barrón-Cedeño et al., 2013). The Greedy string tiling (GST) is used for substring matching in both texts and software plagiarism detection (Clough et al., 2002; Nawab et al., 2010), it involves searching for substrings (in a pair of texts) and merging neighbouring substrings into tiles (long text sequences). The GST is effective in detecting certain modifications in texts such as block move (reordering/transposition), spelling errors, inflectional changes (*with intention/intentionally*), however finding the right substring size is a problem that has resulted in poor performance in some studies (Nawab et al., 2011; Jayapal, 2012).

Virtually all the methods described in this review are limited in some ways; hence in this work we try to optimise the detection of paraphrases by proposing effective methods that target the most common techniques used in paraphrasing texts, and combining the methods into a paraphrase detection model.

## 4  Proposed Methods

This section describes the methods proposed in this paper for detecting paraphrase plagiarism. Each method is a dimension of intertextual similarity designed to detect a particular paraphrase technique, and together they form a paraphrase detection model that detects a wide range of paraphrase plagiarism cases. To ensure that the model generalises, similarity measurement is carried out on a sentence level close to the paraphrase fragments. Sentence similarity across corpora does not vary as much as passage or document level similarity. We described our proposed model below.

### 4.1  The Proposed Paraphrase Retrieval Model

The model takes in a pair of suspect and source text passages, splits them into sentences and pre-processes the sentences using standard NLP pre-processing techniques (Ceska and Fox, 2009; Chong et al., 2011) as follows; it tokenises each sentence into words, normalises the words to lowercase, removes stopwords and diacritics (accent) and stems the remaining words to their root form. Each pre-processed sentence in the suspect passage is compared with sentences in the source passage for similarity using methods for measuring semantic similarity, syntactic similarity (reordering) and insertion/deletion. Sentences with similarity scores below predefined thresholds are discarded. The similarity scores for the sentences are averaged for each method and passed onto a machine learning classifier that classifies the suspect passage as paraphrased or not. We described the methods used in the model below.

### 4.1.1 Method for Measuring Semantic Similarity (Detecting Lexical Substitutions)

Lexical substitutions (synonym replacements) occur when words or short phrases are used to replace other words or short phrases in a passage without changing the meaning of the passage.

The proposed method involves detecting lexical substitutions (synonyms) and co-occurrences between suspect and source sentences (word level semantic similarity), and aggregating the word level similarity into sentence level similarity using a similarity function. To detect lexical substitutions, a combination of query expansion and word pair similarity measurement using WordNet and the word2vec word embedding model is used. Query expansion is used to generate synonyms for each query word in a suspect sentence; the word2vec model transforms the synonyms into word vectors and compares them with word-vectors of the source sentence using cosine similarity. Vector comparisons with similarity scores that exceed a predefined threshold are retrieved. Query words without synonyms (in WordNet) are transformed into word vectors and compared with word vectors of the source sentence using word2vec. When no comparisons result in a similarity score that exceeds the predefined threshold, the Resnik IC measure is used to compare the query word with words in the source sentence, and word-pairs with similarity scores above a predefined threshold are retrieved. The detected words are counted and normalised by the length of the suspect sentence (containment) to give the similarity score of the suspect and source sentences.

We used a pre-trained word2vec model based on Mikolov et al., (2013); it contains 3million words built using Google news articles with a vocabulary size of 100billion words, and a skip-gram model; words and their contexts were transformed into a-300 dimensional vectors (hot vectors) and trained using deep neural network. The steps used in this work to compute the semantic similarity between pairs of sentences are:

1. Expand query word; query WordNet to generate synonyms.
2. Match source words with the synonyms, if match is found, remove query and source words, go to the next query word and repeat the process. This step captures easy to detect cases of word replacements.
3. Else: use word2vec model to transform the generated synonyms and words in the source sentence into word-vectors and compare using cosine similarity. Return match with maximum similarity above threshold; this step uses captures difficult to detect cases of word replacement using (WSD). We used Gensim to implement this model.
4. Else: compare query word with words in source sentence using WordNet Resnik IC and retrieve word pairs with similarity scores above a predefined threshold.
5. Repeat the above steps for all query words in the suspect sentence and normalise the count of the retrieved matches by the suspect sentence length to get the semantic similarity score of the pair. This step essentially measures the semantic similarity of the sentences.

$$semantic\,sim(sp, sr) = \frac{count(lexical\,substitutions)}{len(sp)}$$

### 4.1.2 Method for Measuring Syntactic Similarity (Text Reordering)

Reordering is a paraphrase technique that involves changing the syntactic structure of texts by swapping words or block of words (transposition) without changing the meaning of the passage.

The proposed method for measuring syntactic similarity is similar to Li et al., (2006) method. A word set (union of words) of the suspect and source texts is first formed, the texts are converted into vectors based on the word sequence of the wordset, and the weight of the vector components are the word positions (indices) in the texts. The vectors are then compared using a normalised distance function; we used cosine similarity instead of the normalised distance function proposed by Li et al. See example of how this method works;

Source= {*Mary is the winner of the tournament, and John is the runner up*}

Suspect= {*the winner of the tournament is John, and the runner up is Mary*}

The source and suspect texts are transformed into vectors using their indices as weights;

Source= [1, 2, 3, 4, 5, 6, 7, 8, 9, 10, 11, 12, 13]

Suspect= [13, 12, 1, 2, 3, 4, 5, 8, 7, 6, 9, 10, 11]

$reordering(sp, sr) = cosine(sp, sr)$

Bag of word methods such as the VSM would assign a similarity score of one (exactly similar) to the above sentences because they contain exactly the same words. However, the sentences are not exactly semantically similar. Our proposed method assigns a lower similarity score (0.82) to the pair, which is closer to human judgment, as the similarity score has to reflect the syntactic alterations in the suspect sentence.

### 4.1.3 Method for Measuring Similarity When Insertions/Deletions Are Taking Into Account

Insertion/deletion is a paraphrase technique that involves deleting and inserting words or short phrases in between text in a passage without changing the meaning of the passage.

To measure insertions/deletions, the proposed method involves computing the word level edit distance between the suspect and the source sentences. The edit distance is the minimum edit operations (insert, delete and replace) required to transform a suspect sentence into a source sentence. The edit distance is normalised by the length of the longer sentence, the resultant value is then subtracted from one to give the similarity score of the sentences when insertions/deletions are taking into consideration. The score is in the range of 0 and 1.

- Compute edit distance between source and suspect sentences.
- Divide edit distance by the length of the longer sentence and subtract the value from one.

$$insert/delete(sp,sr) = 1 - \frac{editdistance(sp,sr)}{\max(len(sp,sr))}$$

## 5 Experiments

This section describes the implementation and evaluation of the proposed methods using the Crowd sourcing (Burrows et al., 2012) and the Cloughs and Stevenson (Clough and Stevenson , 2011) corpora, both of which contain simulated paraphrased plagiarised texts with formations that differ from verbatim (cut and paste) and non-paraphrased texts. We describe the corpora below.

### 5.1 Crowd Sourcing Paraphrased Corpus

The crowd sourcing corpus (Burrows et al., 2012) contains 7,859 pairs of passages, of which 4,067 are paraphrased and the remaining 3792 pairs are non-paraphrased. The passages were derived from crowd sourcing, and altered using one or more of the following techniques; synonym replacement, word or phrase reordering, insertion/deletion, inflectional changes of texts etc. The corpus comes with ground-truth that contains details about whether a pair is paraphrased or not.

### 5.2 Cloughs and Stevenson Corpus of Short Plagiarised Answers

The Clough and Stevenson (2011) corpus contains 100 text passages, of which 5 are questions, and the remaining 95 are responses to the questions. The corpus was created by issuing questions to students (respondents) with instructions on how to answer the questions. The questions are original text taken from Wikipedia, some of the respondents were asked to copy and paste their answers, some were asked to paraphrase the original Wikipedia texts lightly and others heavily. The collective responses for all five questions resulted in four categories of plagiarism namely; cut and paste (verbatim copy), light paraphrased, heavy paraphrased and non-plagiarised. The corpus comes with ground-truth that contains the categories in which each response belongs according to human judgment. To use this corpus, we combined the heavy and light paraphrase categories into one category, and run experiments to detect them.

### 5.3 Evaluation of Methods on the Crowd Paraphrase Corpus

We first implemented the semantic, syntactic (reordering) and insert/delete similarity measurement methods in order to determine their relative performances. We made various combinations of these methods (semantic, syntactic and insert/delete) using a machine learning classifier and evaluated their performances. The combinations were carried out using similarity scores from the methods as input features to train and classify text passages as paraphrased or not. Weka 3.8 toolkit with the following classifiers were used in the experiments in order to determine the best performing classifier to use; K-Nearest Neighbour (KNN), Naïve Bayes (NB), and the multi-layer perceptron (MLP). Besides the MLP, these classifiers were outstanding in Burrows et al., (2012) and Bar et al., (2012). A 10-fold cross-validation was used to ensure the methods are generalizable. We then implemented the baseline.

*Baseline*: the greedy string tilling (GST) was used as baseline because it emerged as one of the best performing methods on the P4P paraphrase corpus. The baseline was implemented with the parameters used in Nawab et al., (2010); we normalised each pair of passages (suspect and source) into lower alphabetical case and transform the suspect passage into substrings of five (5) characters in length. We searched for matching substrings in the source passage and joined neighboring matched substrings into tiles, and discard tiles less than 10-characters in length. Similarity was measured using the containment measure as implemented in Clough et al., (2002); length of matching tiles divided by suspect passage length in characters, and a threshold was used to detect paraphrased passages.

*Evaluation Metrics:* The evaluation metrics used in this study are standard in IR and includes precision, recall and f1-measure. We defined precision as the proportion of retrieved passages that are paraphrased, recall as the proportion of paraphrased passages retrieved, and the f1-measure as the harmonic mean of precision and recall.

### 5.3.1  Results and Analysis on the Crowd Paraphrase Corpus

The results obtained from the evaluation on the crowd paraphrased corpus are presented and analysed below.

Table 1: Results from the Experiments on the Crowd Paraphrase Corpus

| Methods | Precision | Recall | F-1 | AUC-ROC |
| --- | --- | --- | --- | --- |
| semantic | 0.800 | 0.923 | **0.857** | 0.915 |
| syntactic | 0.796 | 0.918 | 0.853 | 0.868 |
| Insert/delete | 0.798 | 0.871 | 0.833 | 0.897 |
| Sem, syn | 0.793 | 0.939 | 0.860 | 0.928 |
| Sem, ins/del | 0.789 | 0.936 | 0.858 | 0.916 |
| Syn, ins/del | 0.782 | 0.932 | 0.855 | 0.907 |
| **Sem, syn, ins/del** | **0.803** | 0.938 | **0.865** | **0.917** |
| Bär et al., (2012) | | | 0.852 | |
| Burrows et al., (2012) | | | 0.837 | |
| **Baseline(GST)** | 0.768 | 0.922 | 0.838 | 0.887 |

The results in table 1 are the performances obtained when the proposed methods were run individually and when combined, and the performance of the baseline and previous studies.

The results show that the semantic similarity method outperformed (0.857) the syntactic and insert/delete methods (0.853 and 0.833 respectively) when run individually. This is likely due to the combination of the word level semantic similarity methods used, and in particular the word2vec word embedding model that ensures WSD is applied when detecting similar words, which may have improved the recall. In terms of how well the paraphrased passages are separated from non-paraphrased ones (cost sensitive analysis), the results show a better separation for the insert/delete method than the syntactic method as seen from their AUC (area under ROC) scores. The results also show improvement in performance when the methods were combined. The best performing combination was observed when the semantic, syntactic and insert/delete methods were combined (0.865) using a KNN classifier. The performance of the KNN classifier is consistent with Burrows et al., (2012) study where it emerged as best performing. However when considering AUC, which is a better measure for comparing classifiers, the MLP outperformed the other classifiers. This is likely due to the fact that the KNN is a lazy learner that considers only a few neighboring datapoints when classifying items, while the MLP takes into consideration every datapoints in order to find the best classification boundary between items of different classes.

A closer look at the detected passages revealed that some methods were able to detect paraphrased passages that others could not. This unique ability of some methods to be effective in areas where others are not explains the improved performance obtained from the combination of the methods.

Table 2: Performance from the Classifiers Used in Combining the Methods

| Methods | Precision | Recall | F-1 | AUC-ROC |
|---|---|---|---|---|
| KNN | 0.803 | 0.938 | 0.865 | 0.917 |
| MLP | 0.817 | 0.910 | 0.861 | 0.925 |
| Naïve Bayes | 0.781 | 0.946 | 0.855 | 0.907 |

The results in Table 2 are the performance obtained from the classifiers that were tested in the combination experiments.

In comparison with the baseline and previous studies, the results show that our proposed model (the methods combined) outperformed (F-1: 0.865) the baseline (0.838). The other methods did outperform the baseline as well, except for the insert/delete method. Details of the classification errors of the proposed model and the baseline could be seen in the confusion matrixes in tables 3 and 4 which show classification errors of 0.15 and 0.185 respectively. The error analysis results indicate that the baseline method misclassifies about 1.23 times more than the proposed model.

Table 3: Confusion for the Proposed Model on the crowd paraphrase corpus

|  | Paraphrase | Non |
|---|---|---|
| Para | 3815 | **252** |
| Non | 934 | 2858 |

The result in table 4 shows that out of the 4067 paraphrased passages in the corpus, 252 were misclassified by the proposed model as non-paraphrased.

Table 4: Confusion Matrix for the Baseline on the crowd paraphrase corpus

|  | Paraphrase | Non |
|---|---|---|
| Para | 3748 | **319** |
| Non | 1133 | 2659 |

The result in table 4 shows that out of the 4067 paraphrased passages in the corpus, 319 were misclassified by the baseline method.

Statistical analysis using AUC shows higher AUC value (0.917) for our proposed model, which indicates a better separation of the categories relative to the baseline (0.887). The results also show that the proposed model outperformed previous studies by Bar et al., (2012) (0.852) and Burrows et al., (2012) (0.837) on the crowd paraphrased corpus.

## 5.4 Evaluation of Methods on the Cloughs and Stevenson Corpus

The Cloughs and Stevenson corpus was used as an unknown data set to generalize the methods. We used the parameters (e.g. thresholds) obtained from the previous experiments on the Crowd paraphrase corpus in the generalisation experiments[1]. We evaluated the methods individually, as well as when combined, and we also evaluated the baseline method.

### 5.4.1 Results and Analysis on the Clough and Stevenson Corpus

The results obtained from the Cloughs and Stevenson corpus are presented and analysed below.

Table 5: Results on the Clough and Stevenson corpus

| Methods | Precision% | Recall% | F-1% |
| --- | --- | --- | --- |
| semantic | 78.7234 | 97.3684 | 87.0588 |
| Insert/delete | 75.5102 | 97.3684 | 85.0575 |
| syntactic | 84.2105 | 84.2105 | 84.2105 |
| **Combined** | **87.5** | 92.105 | **89.746** |
| Bär et al.,(2012) | 89.189 | 86.842 | 87.998 |
| **GST (baseline)** | 82.9268 | 89.4737 | 86.0759 |

Table 5 contains the results obtained from the experiments on the Cloughs and Stevenson corpus. The results show that the semantic similarity method outperformed the syntactic and insert/delete methods, which is consistent with the results obtained on the crowd paraphrase corpus. A further increase in performance was observed when the combination of the methods was used for paraphrase detection, which again can be attributed to the combination of the similarity characteristics across the three similarity dimensions (semantic, syntactic and insert/delete).

---

[1] To ensure that the parameters work optimally on both corpora, while developing the methods on the Crowd paraphrase corpora, the parameters were tested on the Clough and Stevenson corpus until a common parameter set that work on both corpora was realised.

The performance difference was however not much, which is likely due to the size of the corpus and errors in the annotations. A close look at some of the text passages revealed that the classification boundary is not clear for some, given the closeness in their similarity scores; this means that some paraphrased passages could easily be classified as non-paraphrased, and vice versa. Details of the classification errors of the combined measure could be seen in the confusion matrix in table 6 which shows that 3 out of the 38 paraphrased passages were misclassified.

Table 6: Confusion Matrix for the Proposed Model on the Clough and Stevenson Corpus

|      | Paraphrase | Non |
|------|------------|-----|
| Para | **35**     | 3   |
| Non  | 5          | 52  |

The confusion matrix in table 6 shows that the misclassification rate of the proposed model on the Clough Stevenson Corpus is 0.084.

In comparison to the baseline and previous studies, the results show that our proposed paraphrased detection method (89.746 %) outperformed the GST baseline (86.0759%), and previous study by Bar et al., (2012) (87.998%) on the Clough and Stevenson corpus. While the size of the corpus may be small, these results revealed insight as to how well the proposed method generalises. The likely reason the methods were able to generalise well is because we carried out comparison on a sentence level, which reduces the variability in similarity scores across the corpora.

## 6 Conclusion

While there are different techniques used in paraphrasing texts, some are more common than others. In this paper we proposed methods for detecting the most common techniques used in paraphrasing texts and combined the methods into a paraphrase detection model. The rationale behind our approach is that if textual similarity characteristics are combined across a number of dimensions that includes the most common techniques used in paraphrasing texts, such a combination could be effectively used to detect most cases of paraphrase plagiarism. We evaluated our approach and obtained results that outperformed a standard baseline and previous studies. The results confirm that the best approach for addressing the paraphrase detection problem is to develop methods for detecting the most common paraphrase plagiarism techniques and combine the methods. Future work will include an evaluation of the proposed methods on a much larger dataset, expansion of the paraphrase detection model to address other textual modifications such as spelling errors.